\documentclass[aps,pre,twocolumn,superscriptaddress,10pt]{revtex4}
\usepackage[dvips]{graphics}
\usepackage{graphicx}
\usepackage{amsfonts}
\usepackage{amssymb}
\usepackage{amsmath}
\usepackage{subfigure}

\begin{document}

\title{Cerenkov-like radiation in a binary Schr{\"o}dinger flow past an obstacle}
\author{H.\ Susanto}
\affiliation{Department of Mathematics and Statistics, University of Massachusetts,
Amherst MA 01003-4515, USA}
%\email{susanto@math.umass.edu}
\author{P.G.\ Kevrekidis}
\affiliation{Department of Mathematics and Statistics, University of Massachusetts,
Amherst MA 01003-4515, USA}
%\email{kevrekid@math.umass.edu}
\author{R.\ Carretero-Gonz\'alez}
\affiliation{Nonlinear Dynamical Systems Group, Department of Mathematics and Statistics,
and Computational Science Research Center,
San Diego State University, San Diego CA, 92182-7720, USA}
%\homepage{http://nlds.sdsu.edu/}
\author{B.A.\ Malomed}
\affiliation{Department of Interdisciplinary Studies, Faculty of Engineering, Tel Aviv
University, Tel Aviv 69978, Israel}
%\email{malomed@eng.tau.ac.il}
\author{D.J.\ Frantzeskakis}
\affiliation{Department of Physics, University of Athens, Panepistimiopolis, Zografos,
Athens 15784, Greece }
\author{A.R. Bishop}
\affiliation{Theoretical Division and Center for
Nonlinear Studies, Los
Alamos National Laboratory, Los Alamos, New Mexico 87545, USA}

%\email{dfrantz@phys.uoa.gr}

\begin{abstract}
We consider the dynamics of two coupled miscible Bose-Einstein
condensates, when an obstacle is dragged through them. The existence of
two different speeds of sound provides
%indicates
%underscores
the possibility for
three dynamical regimes: when both components are subcritical, we do not observe
nucleation of coherent structures; when both components are supercritical
they both form dark solitons in one dimension (1D)
and vortices or rotating vortex dipoles in
two dimensions (2D);
in the intermediate regime, we observe
the nucleation of a structure in the form of a dark-antidark soliton in 1D;
%one dimension, featuring a bright soliton in a nonzero background in the subcritical component;
the
%two-dimensional
2D analog of such a structure, a vortex-lump, is also observed.
\end{abstract}

\maketitle

{\it Introduction.} In the past few years, there has been an
%continuously
increasing number of studies of multi-component
Hamiltonian systems. This has been triggered primarily by the development
of theoretical and experimental results in coupled atomic Bose-Einstein
condensates (BECs) \cite{BEC},
%in atomic physics,
and of coupled
%systems in nonlinear optics
nonlinear optical systems (where the coupling can be, e.g., between
different polarizations of light or different
%propagation
frequencies)
%among other possibilities)
\cite{kivshar}. These have, in turn, motivated
detailed mathematical investigations
of such coupled systems, typically described by nonlinear Schr{\"o}dinger (NLS)
equations \cite{trubatch}.
In the setting of BECs, that will be the
%focal point
primary focus of this study,
mixtures of different spin states of $^{87}$Rb \cite{myatt}
%,cornell_recent}
and $^{23}$Na
%mixed condensates \cite{dsh,dsh1}
\cite{dsh1},
%have been experimentally reported. There have also been efforts to create
as well as two-component
BECs with different atomic species, such as $^{41}$K--$^{87}$Rb
\cite{KRb} and $^{7}$Li--$^{133}$Cs \cite{LiCs}, have been created in experiments.
%more recently to study coherent patterns in them such as vortex
%lattices \cite{cornell_recent}.
In the same context, a wide variety of theoretical studies
have examined
%features such as
ground-state solutions
\cite{shenoy}
and small-amplitude excitations \cite{excit}, as well as
%additional studies include
the formation of other nonlinear structures
%more exotic possibilities
such as
domain walls \cite{Marek}, one-dimensional (1D) bound dark-dark
%\cite{obsantos}
and dark-bright
%\cite{anglin}
soliton complexes \cite{obsantos},
%as well as
spatially periodic states \cite{decon}, vortex dipoles
\cite{kasamatsu}, vortex rings and slaved waves \cite{berloff},
%and even
coupled vortex lattices \cite{mullerho}, and so on.

%On the other hand,
At the same time, many theoretical and experimental studies
%a topic that has been under intense investigation both theoretically and experimentally for one-component BECs
%has been that of
deal with the dragging of an ``impurity''
(e.g., a blue-detuned laser beam) through a {\it one-component} condensate.
This setting has been
%appreciated as
%, in fact,
demonstrated to be prototypical
%one
for dark soliton formation in 1D \cite{Hakim,rad}, and for
vortex formation in 2D
%the formation of dark solitons in one dimension \cite{Hakim,rad}, as well as of vortices in two dimensions
\cite{adams1}. These nonlinear waves can be thought of as a type of nonlinear Cerenkov radiation that
is emitted, when the motion of the impurity is supercritical with respect to the local
%effective
speed of sound of the BEC.
Recently, a combined experimental and theoretical study of the
Cerenkov emission of phonons by a
%blue-detuned
laser obstacle was reported \cite{smerzi};
in a different study \cite{el}, it has been shown that
in the case of large obstacles
%it has been shown that
(and for a supersonic flow of the BEC), the Cerenkov cone transforms into a
spatial shock wave consisting of a chain of dark solitons \cite{el}.
In fact, this setting has been particularly relevant for the study
%has developed into a an interesting testbed for the examination
of the breakdown of superfluidity (and emergence of
dissipation) and the associated Landau criterion
%see, e.g., the discussion of
\cite{pavloff}.
%This has also instigated
Indeed, earlier experiments
%experimental studies in the topic
\cite{kett1}
%that
have demonstrated the onset of dissipation, when
%an impurity in the form of
a blue-detuned laser beam moves through the BEC
%condensate
with velocities above a
%certain speed
threshold. We also note in
passing that the appearance of similar effects
(e.g., the backward-propagating Cerenkov radiation) in
%optical media, such as
photonic crystals \cite{science} is yet another
illustration of the interest in this research direction.
%of study.

%Our scope
In the present work we study
%is to examine the overlap between the two above-mentioned directions, by studying
the dragging of a $\delta$-like obstacle in a {\it two-component} superfluid flow.
If the components are assumed
%considered
to be immiscible,
then they will tend to phase-separate and
%then
the problem reverts to its single-component version.
%installment.
For this reason, we
%examine
consider the case of two miscible components,
%. The latter
which is particularly interesting due to the existence of two distinct
``speeds of sound''.
%which we compute.
%We then examine the phenomenology when dragging a $\delta$-like impurity through the binary condensate.
In this setting, we find two critical speeds $0<v_{c}^{(1)}<v_{c}^{(2)}$. For
$v<v_c^{(1)}$, we show
%find
that the impurity propagates without
emitting Cerenkov radiation in the form of nonlinear waves.
For $v_{c}^{(1)}<v_{c}^{(2)}<v$, both components are supercritical
and the impurity emits gray solitons (in 1D) propagating downstream
in both components. However, the most interesting regime
is the intermediate one, where one of the components is
supercritical, yet the other is subcritical, leading to the
spontaneous formation of dark-antidark solitary waves
previously predicted (in a stationary form) in \cite{epjd}.
We demonstrate that when the strength of the impurity tends
to zero, the critical speeds tend to the corresponding speeds
of sound, yet we show how they deviate from these values for
finite impurity strengths. We also consider the 2D
%two-dimensional
case, where we also obtain the analog of the dark-antidark state
in the form of a vortex-lump wave.

%Our presentation
The paper is structured as follows. We first present
the theoretical framework, and calculate the critical velocities.
%speeds of sound in the two components.
We then numerically investigate
%examine first
the 1D
%one-dimensional case
(both for untrapped and
%in free space, as well as in
trapped BECs)
%condensates; subsequently, we present numerical results for the
%extend our numerical investigations to the
%two-dimensional
and the 2D case.
%In the last section,
Finally, we summarize our
findings and present our conclusions.

{\it Theoretical Setup.} We consider the following
%system of two
coupled NLS equations, describing a quasi-1D binary BEC \cite{BEC}:
\begin{eqnarray}
%i\psi_{1,t}=-\frac12\psi_{1,xx}+\left(g_{11}|\psi_1|^2
%+g_{12}|\psi_2|^2\right)\psi_1+\left(V_1+V_2\right)\psi_1\label{eq1}\\
%i\psi_{2,t}=-\frac12\psi_{2,xx}+\left(g_{12}|\psi_1|^2+g_{22}|\psi_2|^2\right)\psi_2
%+\left(V_1+V_2\right)\psi_2\label{eq2},
i \partial_{t} \psi_{j} = \left( -\frac{1}{2} \partial_{x}^{2} +
\sum_{k=1}^{2} g_{jk}|\psi_k|^2 + V_{\rm ext} \right) \psi_j,
\label{eq1}
%\\
%\label{eq2}
\end{eqnarray}
where $\psi_{j}$ ($j=1,2$) are the mean-field wavefunctions, and $V_{\rm ext}=V_1+V_2$ is the external
potential, assumed to be composed by a repulsive potential of a blue-detuned
laser beam, $V_1$, and a trapping harmonic potential, $V_2$, i.e.,
%
%\[
%\displaystyle
\begin{equation}
%V_1=A\,\exp\left(-\frac{(x-vt)^2}{2\epsilon^2}\right),
V_1=A\,e^{-{(x-vt)^2}/{2\epsilon^2}},
\quad V_2=\frac12\Omega^2x^2,
\label{potentials}
\end{equation}
%\]
%
where $A$, $\epsilon$ and $v$ are, respectively, the strength, width and velocity of the laser obstacle,
and $\Omega$ the harmonic trap strength. The nonlinearity coefficients are chosen to be
$g_{11}:g_{12}:g_{22}=1.5:1:1.03$.
%, having in mind
Notice that two of them have the ratios that
%they typically do
are typical for $^{87}$Rb \cite{myatt},
while the third is tuned to a different value (so as to ensure miscibility since
the standard value of $g_{11}=0.97$ would lead to immiscibility).
The tuning can be achieved by means of a Feshbach resonance \cite{feshbach}.
%We look for stationary solutions of the form:
%\begin{equation}
%\psi_{1,2}\rightarrow\psi_{1,2}e^{-i\mu_{1,2}t}.
%\end{equation}
Moreover, throughout the paper we use the following parameter values
for our numerical computations: chemical potentials
%$g_{11}=1.5,\,g_{12}=1,\,g_{22}=1.03,\,
$\mu_1=1.2,\,\mu_2=1$,
%and
obstacle width $\epsilon=0.5$, and harmonic trap strength
%When there is a magnetic trap, we use
$\Omega=0.02$. The results
do not change qualitatively for other parameter values.
%reported have been found to be typical also for different values of the above parameters.

The uniform solutions of Eqs.~(\ref{eq1})
%will then
satisfy
\begin{eqnarray}
|\psi_1^{(0)}|^2=\frac{\mu_1 g_{22}-\mu_2 g_{12}}{\Delta}, \,\,
%\label{un1}
%\\
|\psi_2^{(0)}|^2=\frac{\mu_2 g_{11}-\mu_1 g_{12}}{\Delta},
\label{un2}
\end{eqnarray}
where $\Delta=g_{11}g_{22}-g_{12}^2$.
%Going to
Expressing Eqs.~(\ref{eq1}) in the traveling-wave frame (i.e., $x \rightarrow x-v t$) and linearizing
around these uniform states according to $\psi_j=\psi_j^{(0)}+ \psi_j^{(1)}$
%\delta r_j$,
we obtain the equations for the small
%linearization
amplitudes $\psi_j^{(1)}$,
%$r_j$
%
\begin{eqnarray}
\frac{1}{4} \frac{d^2}{dx^2} \psi_j^{(1)}
%r_{j,xx}
=\left(
%g_{jj} \left(\psi_j^{(0)}\right)^2
c_j^2
-v^2\right)\psi_j^{(1)}
+ g_{jk} \psi_j^{(0)} \psi_k^{(0)} \psi_k^{(1)},
%r_k,
\label{lin1}
\end{eqnarray}
with $j,k\in\{1,2\}$, $k \neq j$ and $c_j^2=g_{jj} (\psi_j^{(0)})^2$.
Writing this system
%of two second-order equations
as one of four first-order equations
(for $\Psi_j^{(1)}= \dot{\psi}_j^{(1)} \equiv (d/dx) \psi_j^{(1)}$ and $\psi_j^{(1)}$),
namely,
%(for $q_j=r_{j,x}$ and $r_j$)
%
\begin{equation}
%\frac{d}{dx}
\left(\begin{array}{c} \dot{\Psi}_1^{(1)} \\ \dot{\psi}_1^{(1)} \\ \dot{\Psi}_2^{(1)} \\ \dot{\psi}_2^{(1)}
%q_1 \\ r_1 \\ q_2 \\ r2
\end{array} \right)
= \left(
\begin{array}{cccc}
0& c_1^2 -v^2  & 0 & b \\[0.7ex]
1& 0 & 0 & 0 \\[0.7ex]
0 & b & 0 & c_2^2 -v^2 \\[0.7ex]
0 & 0 & 1 & 0
\end{array}
\right)
\left(\begin{array}{c}
%q_1 \\ r_1 \\ q_2 \\ r2
\Psi_1^{(1)} \\ \psi_1^{(1)} \\ \Psi_2^{(1)} \\ \psi_2^{(1)}
\end{array} \right),
\end{equation}
where
%$c_j^2=g_{jj} (\psi_j^{(0)})^2$ and
$b=g_{12} \psi_1^{(0)} \psi_2^{(0)}$.
Then, the $4\! \times\! 4$ matrix has eigenvalues
$\lambda^2=\tilde{c}^2-v^2 \pm \sqrt{\tilde{c}^4-\Delta
(\psi_1^{(0)} \psi_2^{(0)})^2}$, with
$\tilde{c}^2=(c_1^2+c_2^2)/2$. For stability, we need the
eigenvalues to be real, hence $\lambda^2>0$ implies that
$v<v_{c}^{(1)}<v_c^{(2)}$, where
the critical velocities, corresponding to the two distinct speeds of sound, are given by
$v_{c}^{(1,2)}=\tilde{c}^2 \pm
\sqrt{\tilde{c}^4-\Delta  (\psi_1^{(0)} \psi_2^{(0)})^2}$.
%These critical velocities correspond to the speeds of sound of the two components.
For the parameters mentioned above,
$v_c^{(1)}=0.34393$ and $v_{c}^{(2)}=1.04796$. We expect that
superfluidity will break down when the speed $v$ of the defect
%moving across the condensate
overcomes these speeds; in fact, as argued
in \cite{Hakim,pavloff}, the actual critical point should be
expected to be lower than the above Landau prediction.

%%%%%%%%%%%%%%%%%%%%%%%%%%%%%%%%%%%%%%%%%%%%%%%%%%%%%%%%%%%%%%%%%%%%%%%%%%%%%%%
\begin{figure}[tbh]
\centerline{\includegraphics[width=7cm,height=2.2cm,angle=0,clip]{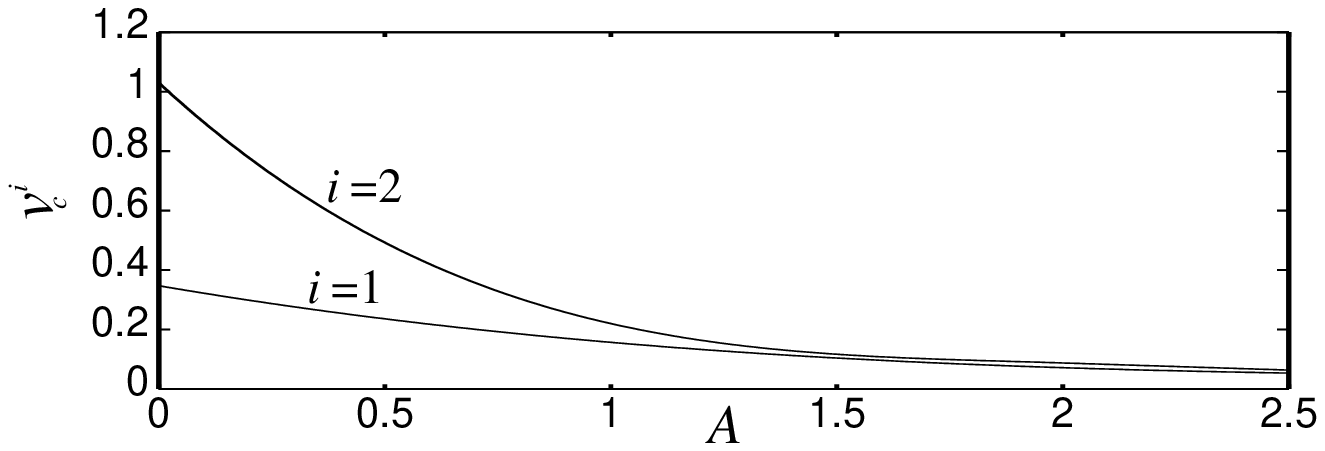}}
\vskip0.1cm
%\centerline{\includegraphics[width=7cm,angle=90,clip]{Fig1bB.eps}}
\centerline{
\includegraphics[width=4.5cm]{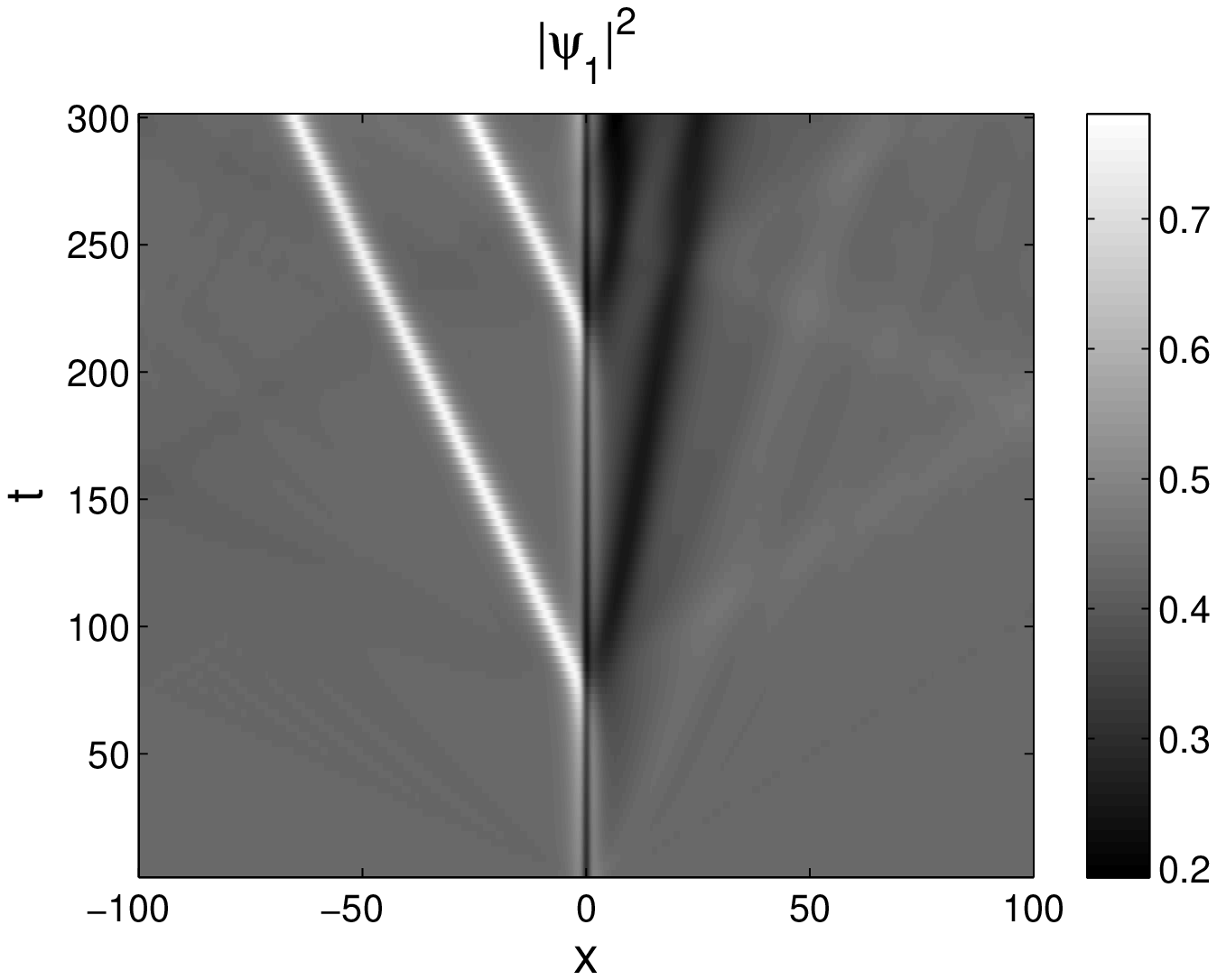}
\includegraphics[width=4.5cm]{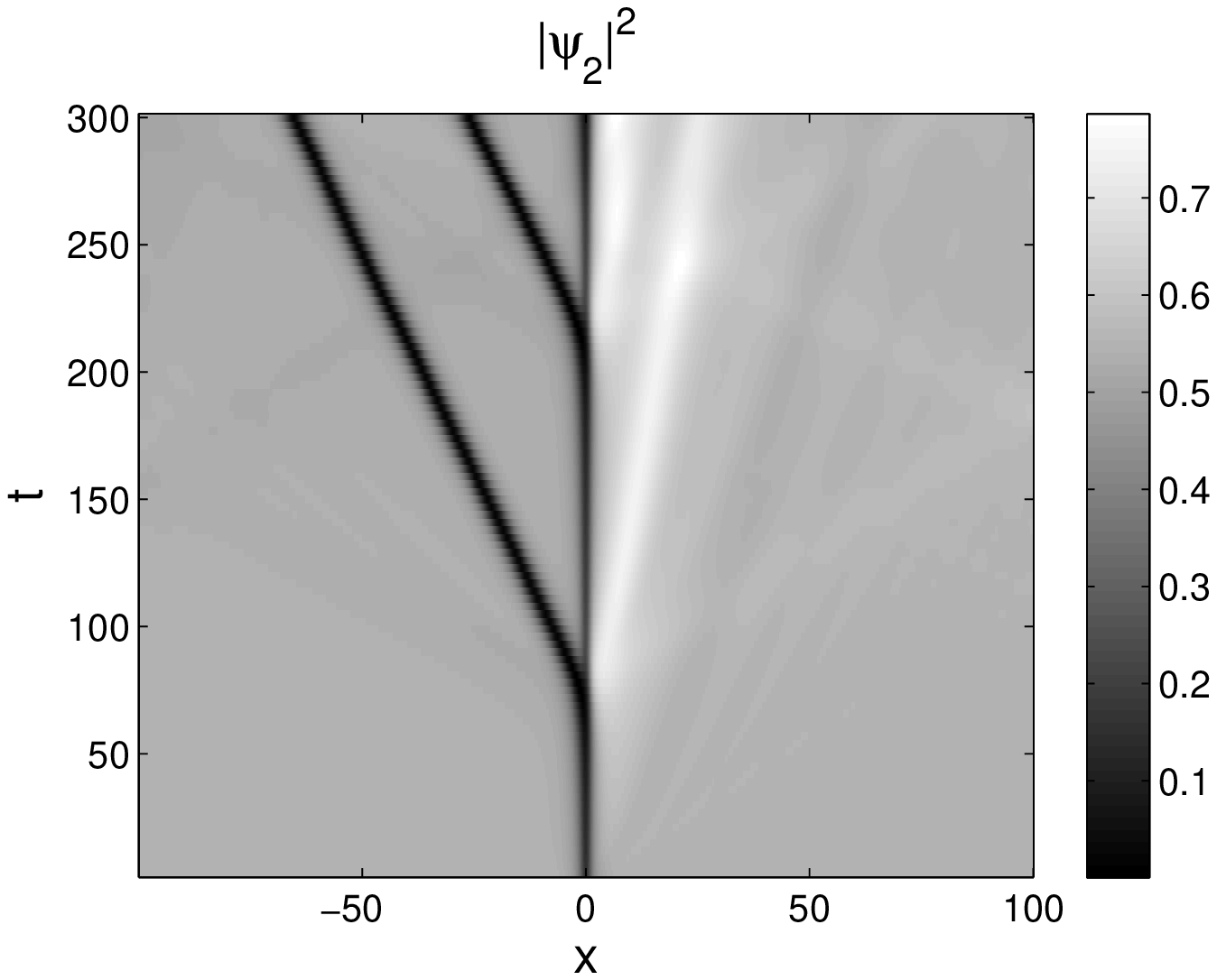}
}
\centerline{
\includegraphics[width=4.5cm]{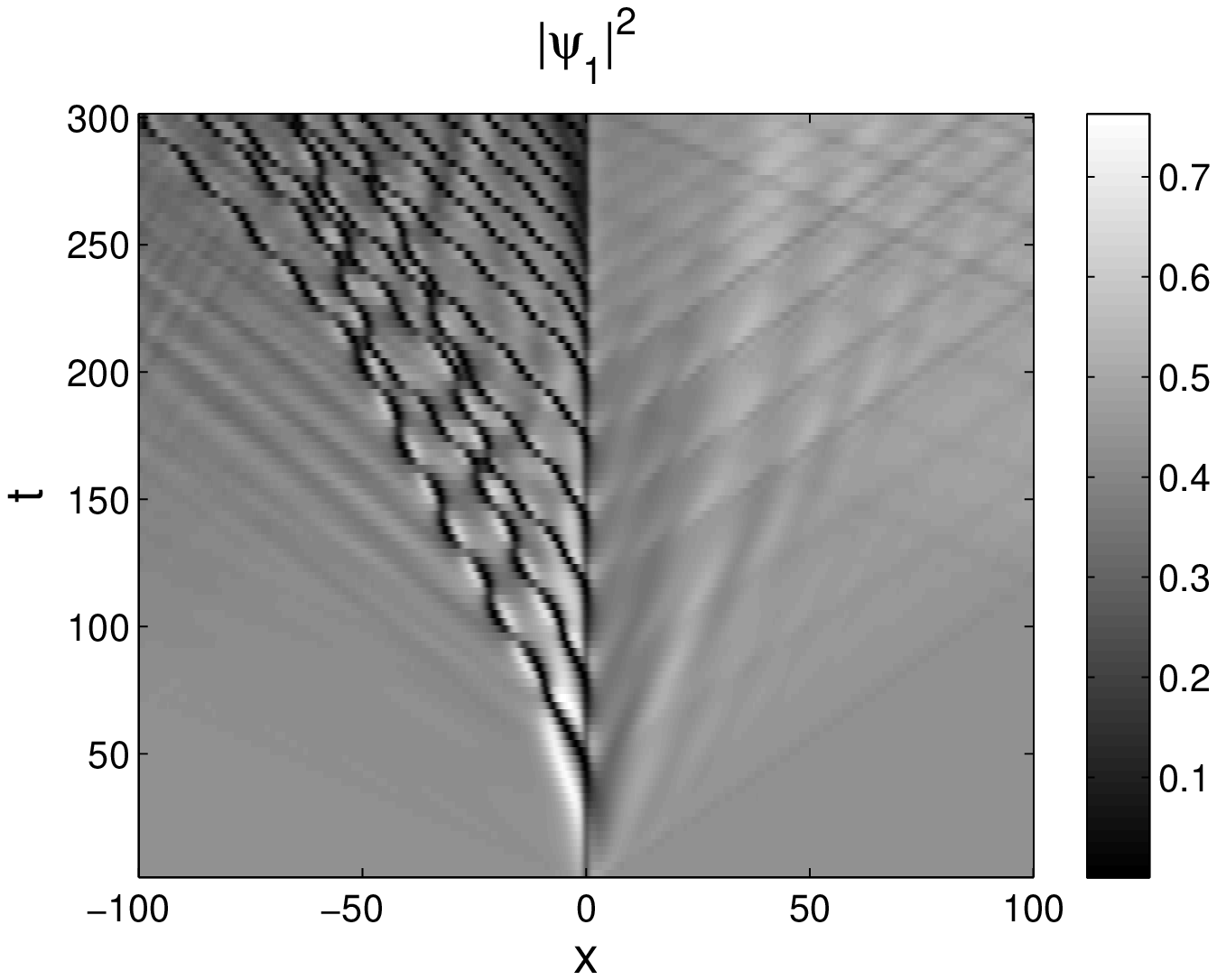}
\includegraphics[width=4.5cm]{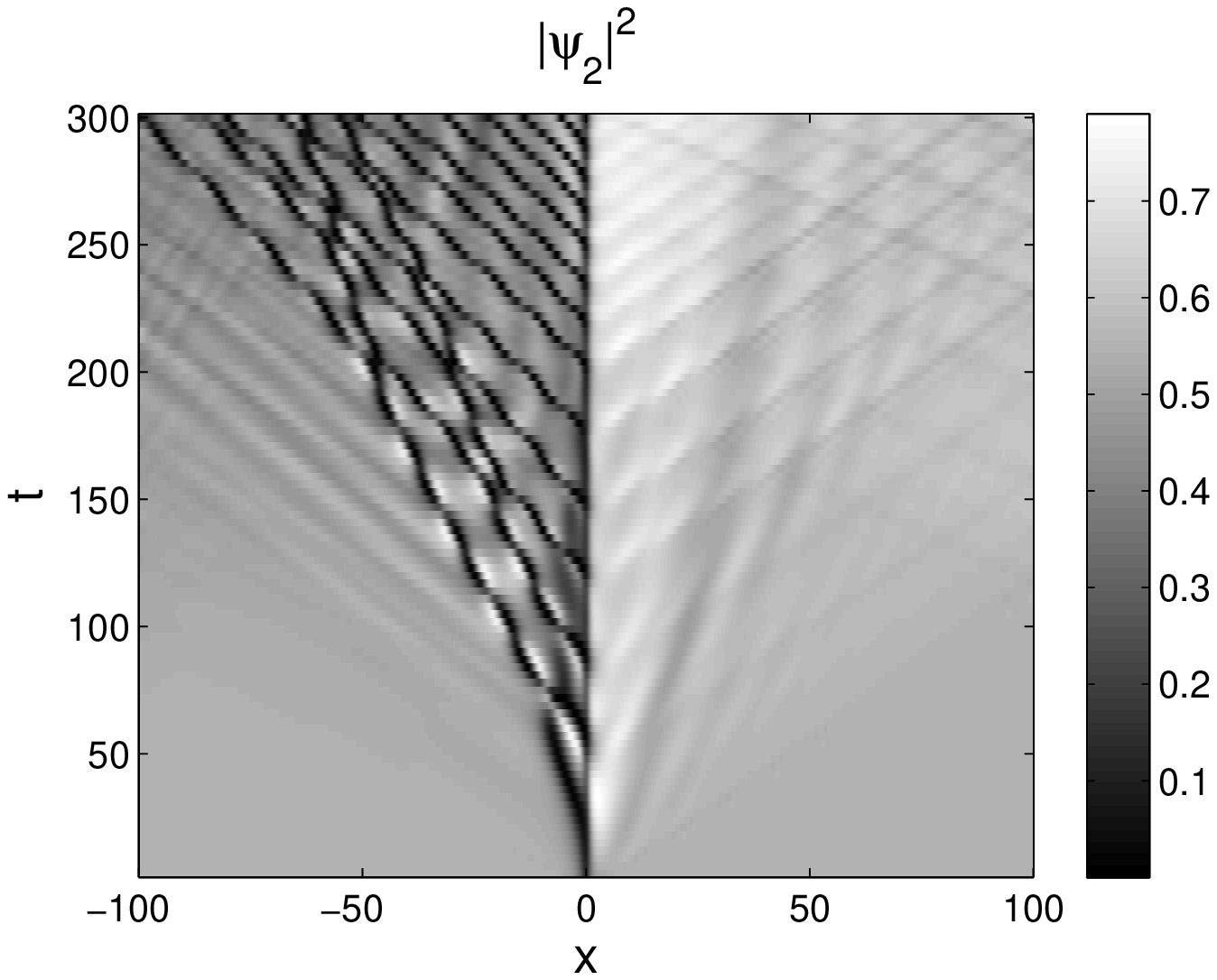}
}
\caption{
%(Color online).
Top panel: ``Stability'' boundaries for static dark soliton pairs:
the critical velocities for the
first and the second component are shown as a function of the impurity
strength $A$.
%, from numerical computations.
Middle panels: Space-time evolution of the density contours for the
two components in moving coordinate frame with velocity $v=0.2$ (the speed
of the impurity); clearly the impurity induces the radiation of dark-antidark
pairs. Bottom panels: similar to the middle panels but with
%the velocity of the obstacle
$v=0.3$, where both dark-antidark dipoles
%(see text)
and dark-dark pairs are emitted.
Impurity parameters: $A=0.9$ and $\epsilon=0.5$.
}
\label{Fig1}
\end{figure}
%%%%%%%%%%%%%%%%%%%%%%%%%%%%%%%%%%%%%%%%%%%%%%%%%%%%%%%%%%%%%%%%%%%%%%%%%%%%%%%

{\it Numerical Results.}
We now turn to the numerical investigation of the above setting.
%and analytical predictions.
In Fig. \ref{Fig1}, we test the theoretical prediction for the existence of two
critical velocities for the dynamical evolution in the two
components. The top panel of the figure shows a relevant
``bifurcation diagram'', where the dependence of the critical
velocities on the ``strength'' $A$ of the
%potential
impurity is numerically evaluated. Note that as the strength of the
impurity tends to zero ($A\rightarrow 0$) one recovers the numerical
values for $v_c^{(1,2)}$ stated above.
The critical velocities are computed by finding (i) the speed
$v_c^{(1)}$ above which, apparently, one component is supercritical
emitting dark solitons, while the other is subcritical emitting
antidark solitons (i.e., bright solitons on a finite background)
that ``accompany'' the dark ones; and (ii) the speed
$v_c^{(2)}$ above which both
components nucleate dark solitons (see also bottom panel
of the figure for a space-time evolution of the density contour plot
for the two components for such a supercritical ---in both components---
velocity $v$).  It is
%worthwhile to note
noteworthy that
such structures had been predicted in a steady form in \cite{epjd},
but, to
%the first best of
our knowledge, this is the first demonstration
of their dynamical nucleation.
%There is one more
A further observation is
worth making
%, in passing,
about the case of $v>v_{c}^{(2)}$. Note
that, especially for early times, the impurity (stationed at $x=0$
%$\xi=0$
in the computations of Fig. \ref{Fig1} performed in the co moving
frame) initially emits structures that
%look
appear more like dark-antidark
{\it dipoles}, i.e., dark-antidark pairs in one-component coupled
with antidark-dark pairs in the other component. This is
%also
again the first manifestation of such structures
%as well;
(to the best of our knowledge);
however, we will make a connection below to
their 2D
%two-dimensional
analog that has been previously proposed \cite{berloff2}.

%%%%%%%%%%%%%%%%%%%%%%%%%%%%%%%%%%%%%%%%%%%%%%%%%%%%%%%%%%%%%%%%%%%%%%%%%%
\begin{figure}[t]
\vspace{-0.5cm}
%\centerline{\includegraphics[height=8.5cm,angle=90,clip]{Fig2B.eps}}
\centerline{
\includegraphics[width=4.5cm]{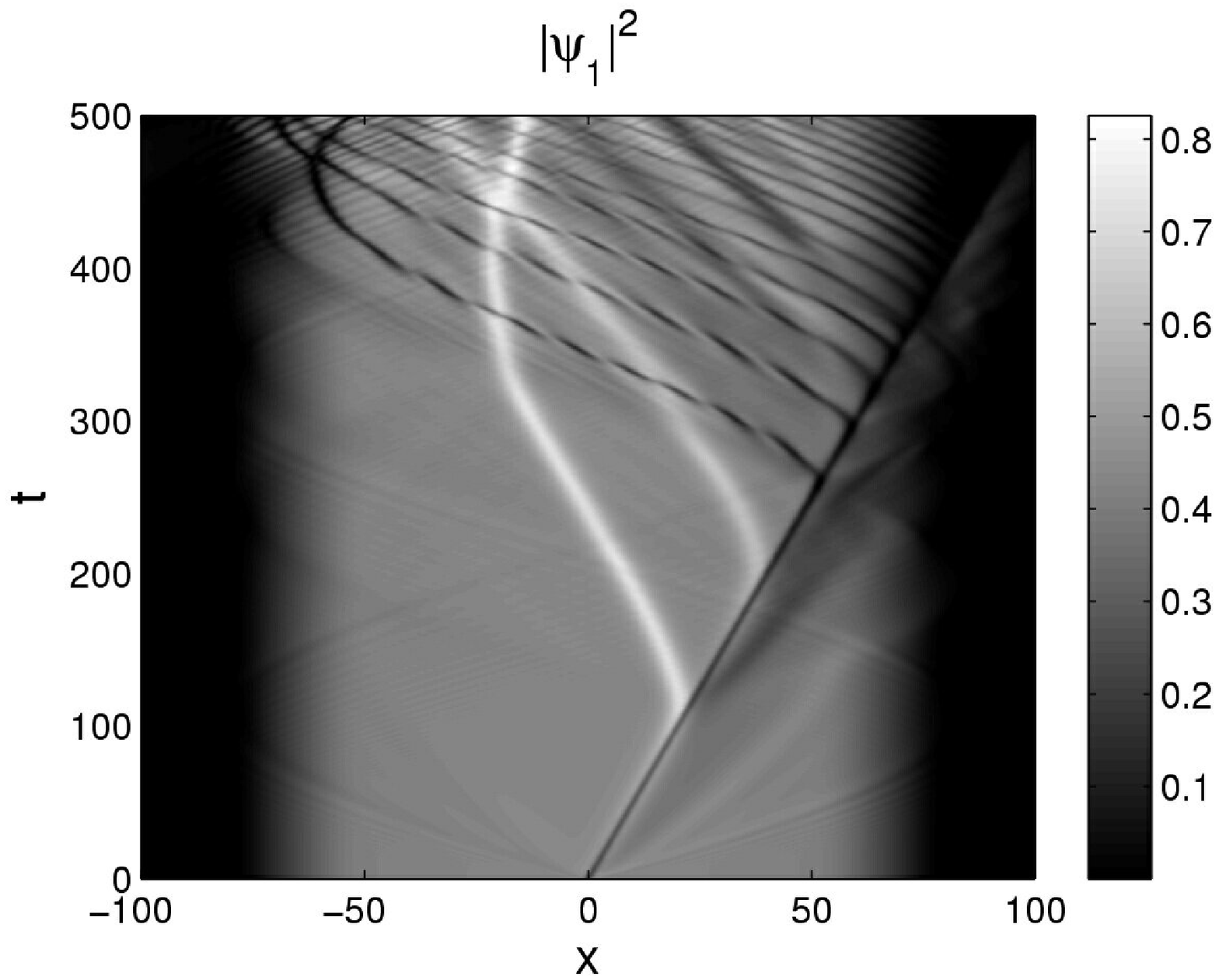}
\includegraphics[width=4.5cm]{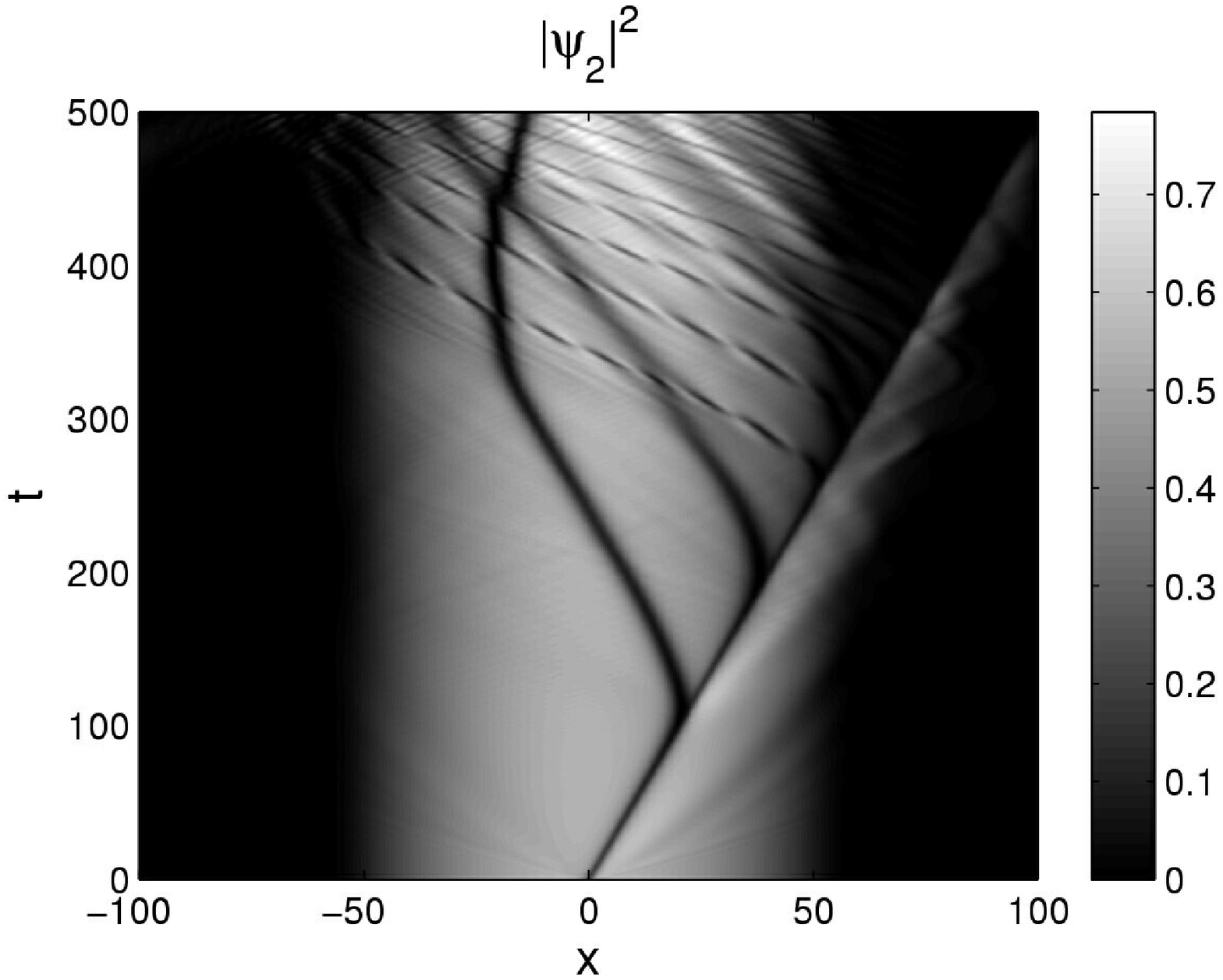}
}
\centerline{
\includegraphics[width=4.5cm]{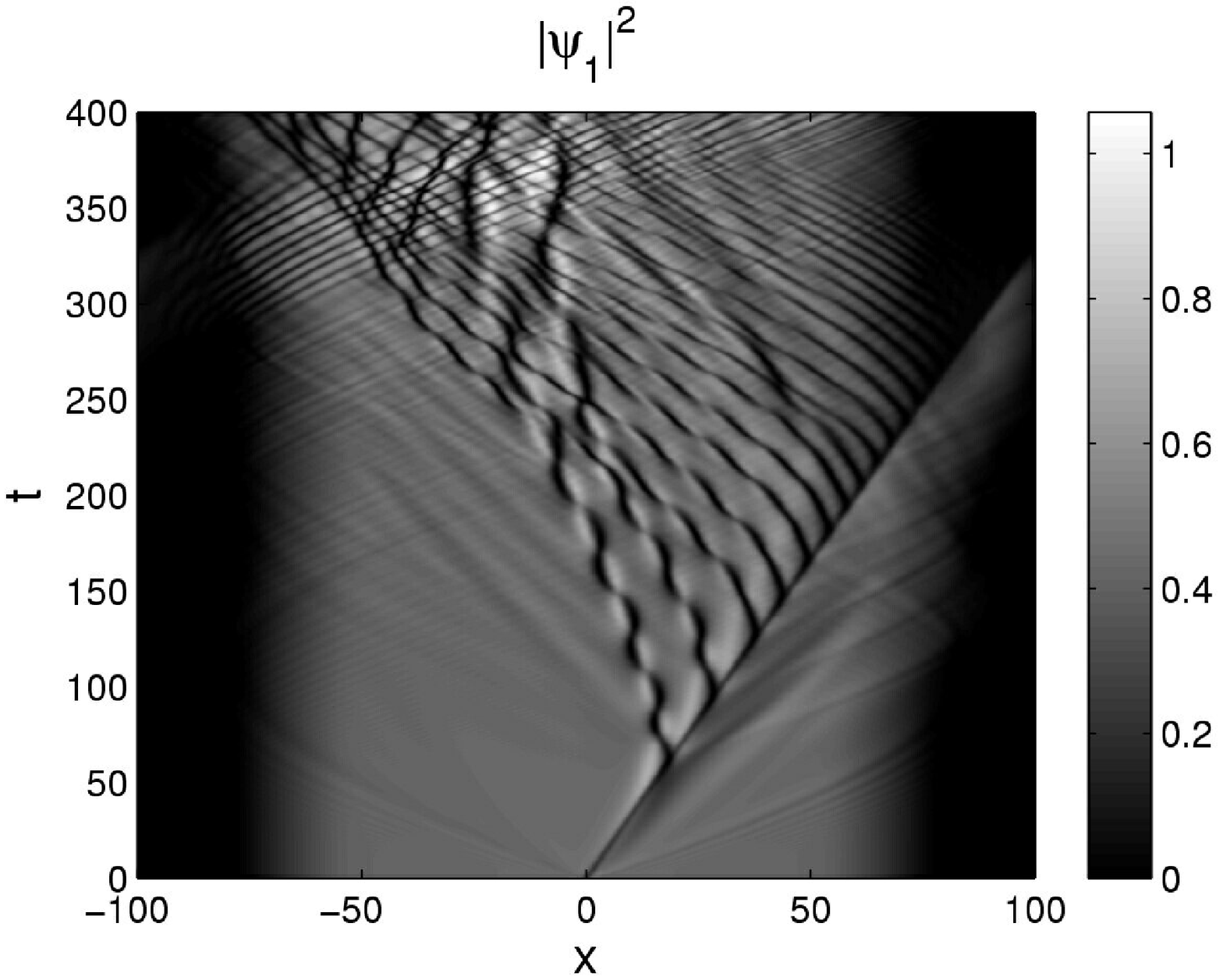}
\includegraphics[width=4.5cm]{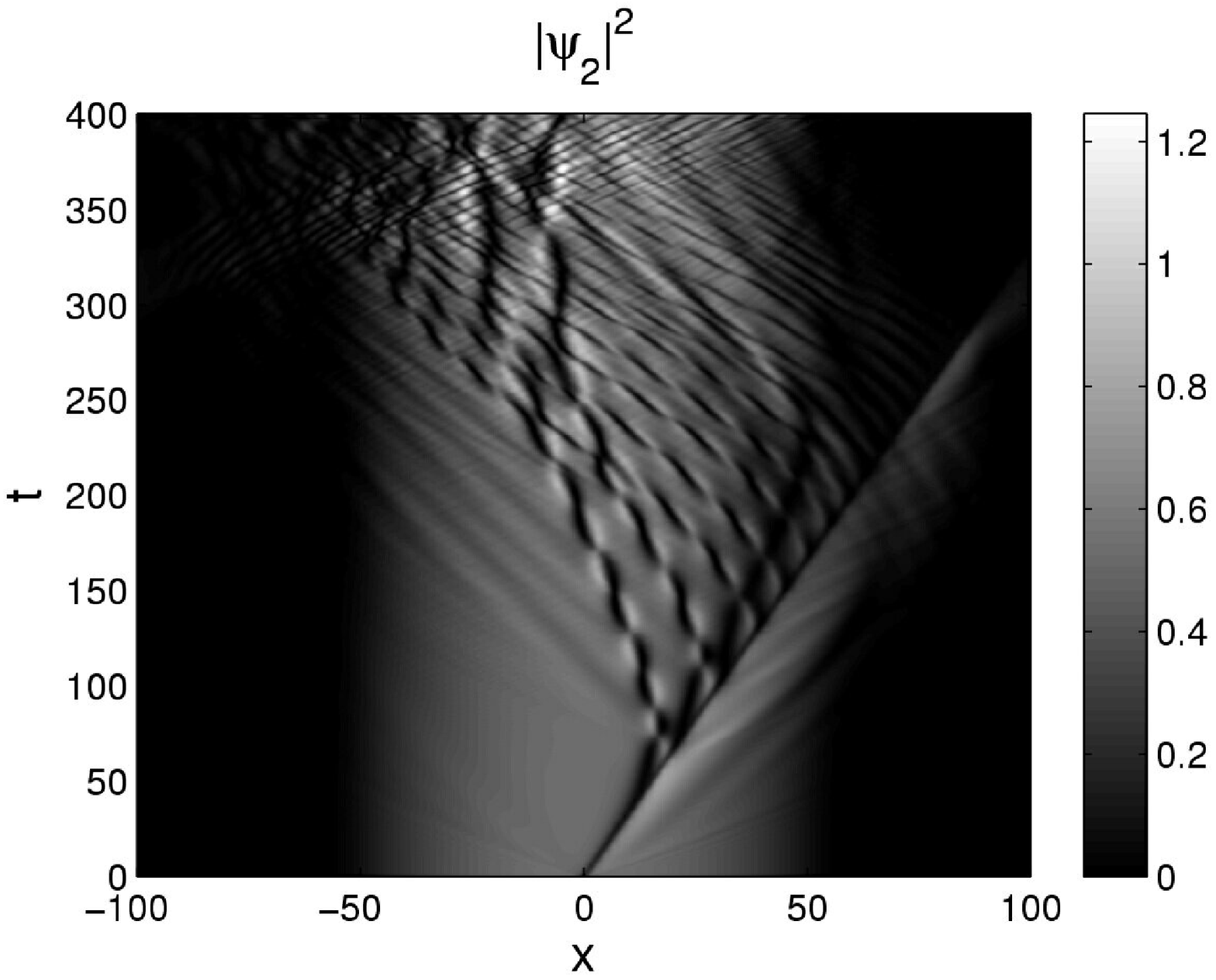}
}
\caption{Space-time contour plots of the components' density,
when the obstacle's
%is moving with
velocity is $v=0.2$ (top panels) and
$v=0.3$ (bottom panels) in the
%presence of a magnetic trap
trapped case
%with
($\Omega=0.02$).
Impurity parameters: $A=0.9$ and $\epsilon=0.5$.
}
\label{Fig2}
\end{figure}
%%%%%%%%%%%%%%%%%%%%%%%%%%%%%%%%%%%%%%%%%%%%%%%%%%%%%%%%%%%%%%%%%%%%%%%%%%
%
%%%%%%%%%%%%%%%%%%%%%%%%%%%%%%%%%%%%%%%%%%%%%%%%%%%%%%%%%%%%%%%%%%%%%%%%%%
\begin{figure}[ht]
%\hskip0.00cm (A) $v = 0.230$
\hskip0.15 cm (A) $v = 0.235$
\hskip2.45 cm (B) $v = 0.350$\\[1.0ex]
\centerline{
\includegraphics[width=4.4cm,angle=0,clip]{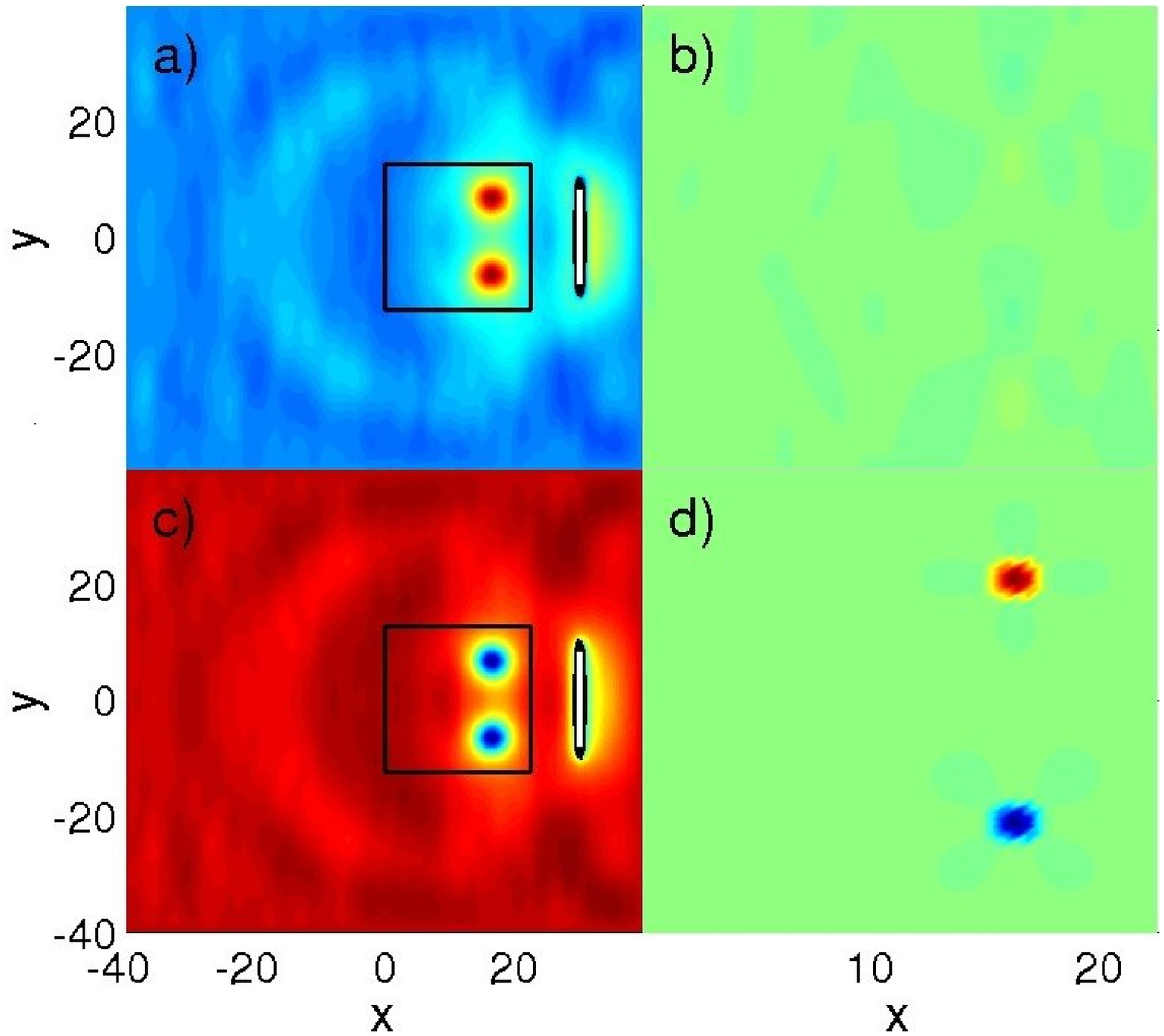}
\includegraphics[width=4.4cm,angle=0,clip]{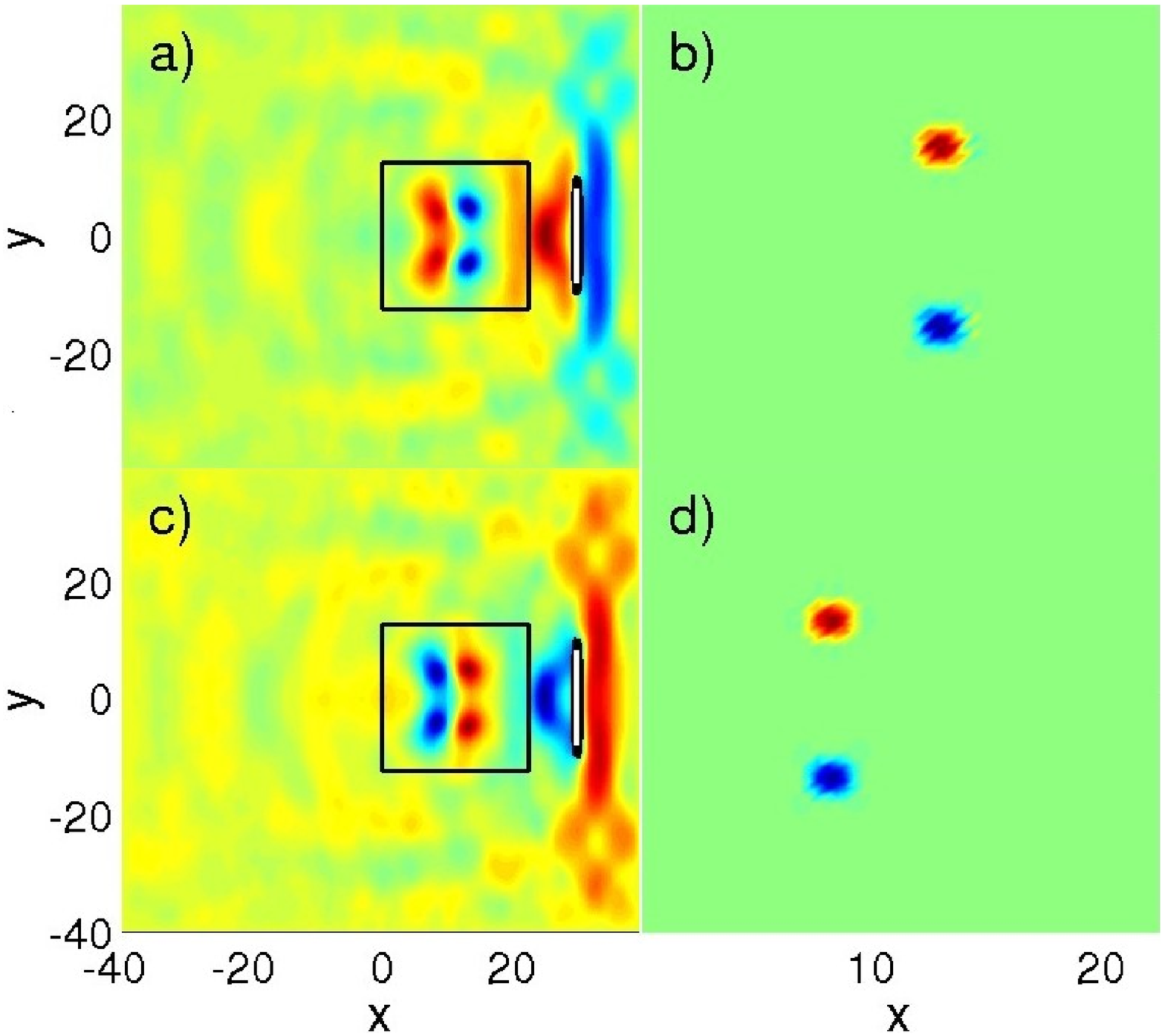}
~~
}
%\medskip
%\medskip
%\hskip0.0cm (C) $v = 0.335$ \hskip6.7 cm (D) $v = 0.350$\\[1.0ex]
%\centerline{
%\includegraphics[width=8cm,angle=0,clip]{w10_v0.335_crop.jpg.ps}
%~~~~
%\includegraphics[width=8cm,angle=0,clip]{w10_v0.350_crop.jpg.ps}
%}
\caption{
(Color online).
Final snapshots after vortex nucleation for different
velocities of a running impurity of size $w=10$ (elongated
vertical bar in panels a and c).
The two depicted cases correspond to:
%(A) $v = 0.230$: $v$ is subsonic for both components.
(A) $v = 0.235$ is subsonic for component $\psi_1$ and
supersonic for component $\psi_2$.
%Velocity is just past the first critical value.
(B) $v = 0.350$ is supersonic for {\em both} components.
The top (bottom) panel corresponds to
component $\psi_1$ ($\psi_2$). Left panels
(a,c) show the square modulus of the solution together with the
moving impurity. The right panels (b,d) show the vorticity $\omega$
in the rectangular area depicted in the left panel counterparts.
Red/blue (top/bottom spots) corresponds to regions
with positive/negative vorticity.
}
\label{fig:2Ca}
\end{figure}
%%%%%%%%%%%%%%%%%%%%%%%%%%%%%%%%%%%%%%%%%%%%%%%%%%%%%%%%%%%%%%%%%%%%%%%%%%

%%%%%%%%%%%%%%%%%%%%%%%%%%%%%%%%%%%%%%%%%%%%%%%%%%%%%%%%%%%%%%%%%%%%%%%%%%
\begin{figure}[tbh]
%\hskip0.7cm (A) $v = 0.350$ \hskip 2.3 cm (B) $v = 0.325$\\[1.0ex]
%\centerline{
%\includegraphics[width=5cm,angle=0,clip]{w15_v0.350_crop.jpg.ps}
%~~~~
%\includegraphics[width=5cm,angle=0,clip]{w20_v0.325_crop.jpg.ps}
%}
%\hskip0.0cm (A) $v = 0.350$ \hskip6.7 cm (B) $v = 0.325$\\[1.0ex]
\centerline{
\includegraphics[width=4.2cm,height=6.6cm,angle=0,clip]{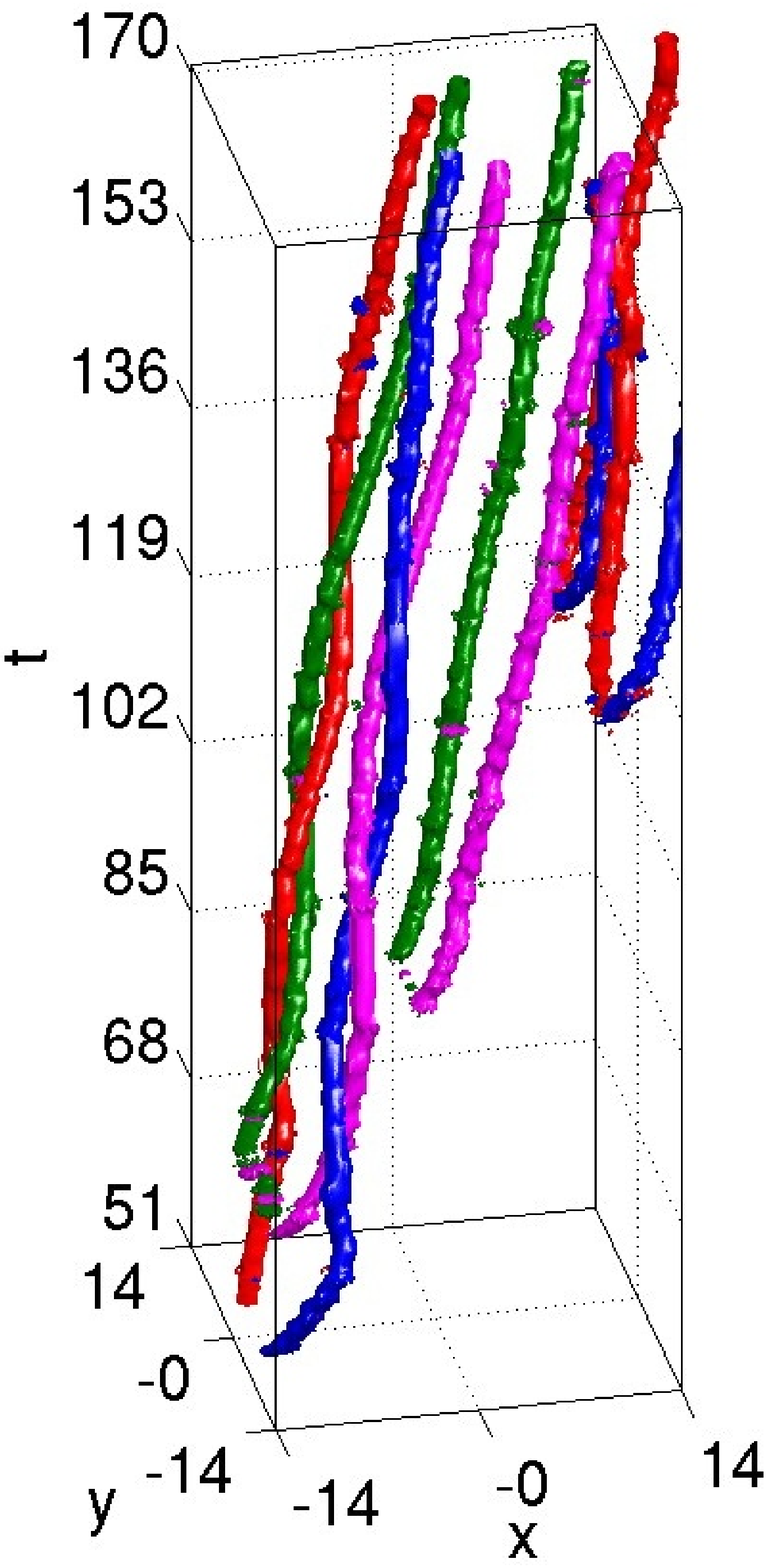}
\includegraphics[width=3.9cm,height=6.6cm,angle=0,clip]{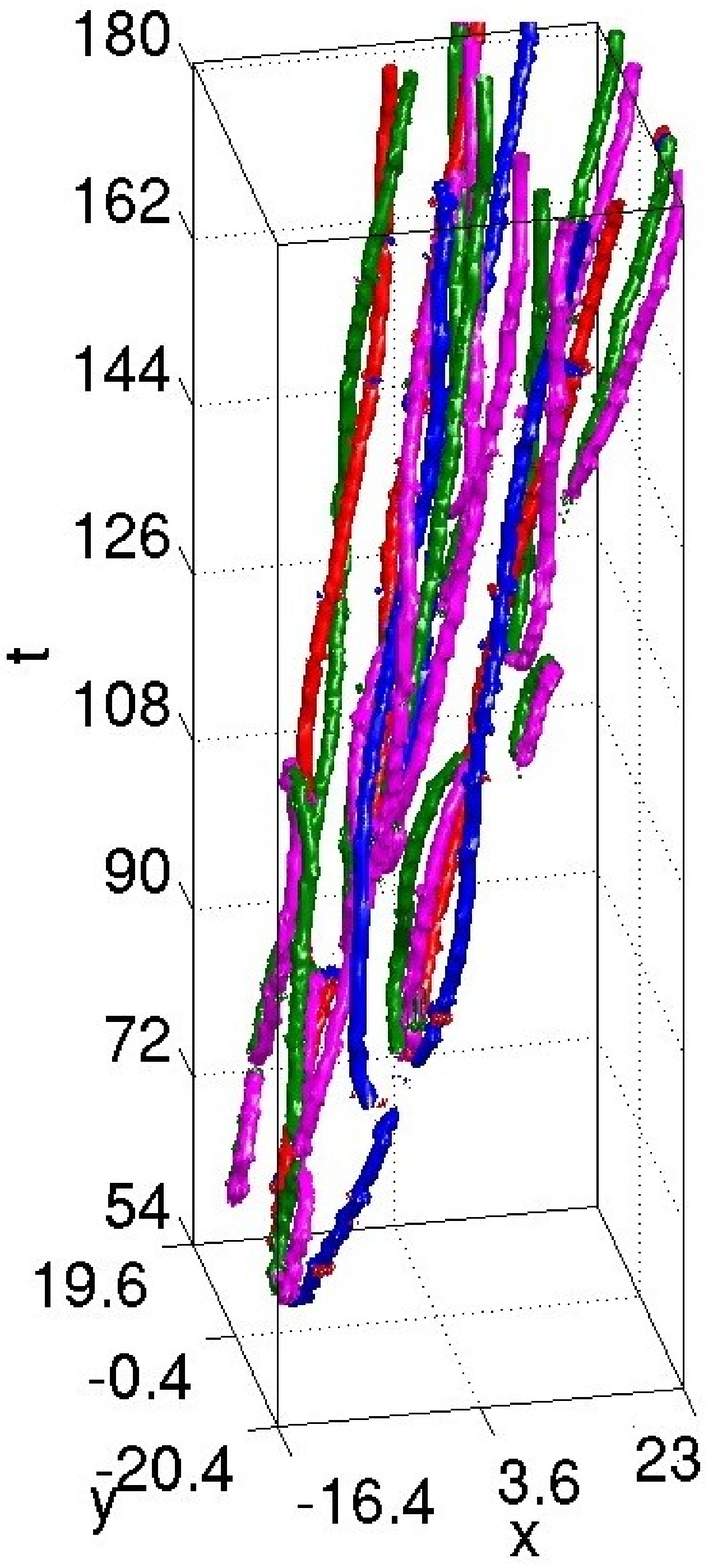}
}
\caption{
(Color online).
Vortex nucleation by a running impurity.
%The first two rows are similar to Fig.~\ref{fig:2Ca} but
%for a wider impurity of width $w=15$ (left) and $w=20$ (right), which
%%Notice how a wider impurity
%is able to nucleate more complex vortex patterns.
The panels depict 3D contour
plots of the vorticity $\omega(x,y,t)$ for the different velocities
and impurity size combinations:
left panel: $w=15$ and $v=0.35$ and
right panel: $w=20$ and $v=0.325$.
Both cases correspond to
impurity velocities that are supersonic for {\em both} components.
The red/blue isosurfaces correspond to positively/negatively
charged vortices in the {\em first} component. We also
superimpose the vorticity isocontours of the {\em second} component
where green/magenta correspond to positive/negative vorticity.
Note how {\it rotating vortex dipole} pairs between the two
components are formed (left-most red-green and blue-magenta intertwined
vorticity lines).
}
\label{fig:2Cb}
\end{figure}
%%%%%%%%%%%%%%%%%%%%%%%%%%%%%%%%%%%%%%%%%%%%%%%%%%%%%%%%%%%%%%%%%%%%%%%%%%

%While Fig. \ref{Fig1} presents the relevant phenomenology in
%free space and in the absence of a magnetic trap, the relevant
The above phenomenology
%will
also persists in the presence of an harmonic
%a magnetic
trap, which is a more realistic setting for magnetically
confined BECs. This is clearly shown in Fig. \ref{Fig2}, where all parameters
are the same as in the corresponding plots of Fig. \ref{Fig1}, but
%in addition a magnetic
incorporating an harmonic trap of frequency $\Omega=0.02$.
%has been applied.

We now turn to the 2D case where the second
spatial derivatives in Eq.~(\ref{eq1})
are substituted by the Laplacian and the impurity potential
is replaced by its 2D counterpart
$V_1(x,y)=(A/4)\exp\left(-{(x-vt)^2}/{2\epsilon^2}\right)
(\tanh(y+w/2)+1)(\tanh(-y+w/2)+1)$,
modeling a light sheet of strength $A$, width $\epsilon$, and size $w$
(see elongated vertical bar in panels a and c of Fig.~\ref{fig:2Ca}).
In our 2D simulations we took $A=0.9$, $\epsilon=0.5$ (i.e.~same
parameters as for the 1D case), and $w=10,15,20$.
Given the similarities of the
trapped and untrapped case in the relevant phenomenology,
we only show the latter here.
In Fig.~\ref{fig:2Ca} we illustrate the two regimes leading to vortex
nucleation (the trivial regime for subcritical velocity in both
components is not shown here):
%, one example
%of each regime discussed previously is presented:
%Panel (A) shows the case
%of $v=0.23$ which is subcritical in both components;
%
(A) $v=0.235$ is subcritical in the first component, but supercritical
in the second, resulting for the latter in a vortex state which is
coupled to a lump (a 2D structure on a finite background)
in the first component. Notice that the presence of vortex states
is clearly illustrated in all the figures contained herein, by
means of the contours of the vorticity
$\omega=\nabla \times v_s$, where $v_s=(\psi^*\nabla \psi -
\psi \nabla \psi^*)/i|\psi|^2$ is the velocity field.
Such structures have been reported
previously for $g_{11}=g_{22}$ in \cite{berloff2}.
(B) $v=0.235$ is supercritical for {\it both}
components. This results in the formation of a dipole
state which contains a vortex-lump pair, coupled to a lump-vortex
pair, in a form similar to the stationary states reported in \cite{kasamatsu}.
%To the best of our knowledge, the spontaneous dynamical nucleation of
%such a dipolar state has not been previously reported.

The last case (supersonic in both components) is examined
in further detail in Fig.~\ref{fig:2Cb}
for different velocities and widths of the quasi-1D obstacle.
In the figure, the actual
spatio-temporal evolution of the vorticity is shown.
This clearly reveals the presence of a
{\it vortex dipole} between the two components; moreover,
this robust type of state appears to be clearly rotating,
as time evolves. Furthermore, it can be noted
that the wider the obstacle, the more complex the
ensuing vortex patterns
%are going to
will be, with multiple
vortex pairs being emitted.

{\it Conclusions.}
%In the present paper,
We have considered
%the case of two miscible components in Bose-Einstein condensates and
the nucleation of coherent structures by a moving obstacle
in two miscible BEC components.
%this setting.
In one spatial dimension, we identified
three different regimes: one without nucleation; one involving
the nucleation of dark-antidark solitons previously predicted
in stationary form in \cite{epjd}; and one producing dark-dark
soliton pairs, as well as dark-antidark dipoles. The critical
points between these regimes were numerically obtained and,
consistently with the corresponding single-component theory,
were shown to approach the Landau criterion for impurity
strength tending to zero; they were systematically lower than
that as this strength increased. It was shown that
similar behavior also occurs in the case of the harmonically trapped
coupled BECs.
%condensates confined by a magnetic trap.
We also examined the same type of behavior in 2D
%two-dimensional
systems. We observed the existence of similar types of regimes, as in the
%one-dimensional
1D case (subcritical in both, supercritical in one,
and supercritical in both). The intermediate regime gave rise
to vortex-lump type structures (also discussed in \cite{berloff2}),
while the supercritical regime
gave the first example of nucleation of vortex-lump dipoles
(obtained in stationary form in \cite{kasamatsu}),
which were actually observed to be rotating as time evolved.
This investigation indicates that there is an interesting
%range
spectrum of dynamical possibilities available in multi-component condensates,
which it would certainly be relevant to explore experimentally.
The recent realization of spinor condensates with more than
two components may provide a fertile ground for further theoretical
investigations in such higher-component settings.

%\vspace{5mm}

{\it Acknowledgments.} PGK acknowledges support from
NSF-DMS-0204585, NSF-DMS-0619492 and NSF-CAREER. RCG and PGK also
acknowledge support from NSF-DMS-0505663. The work of BAM was
partially supported
%in a part,
by the Israel Science Foundation through the
Center-of-Excellence grant No.~8006/03. Work at LANL is performed
under the auspices of the US DoE.

\end{document}